\documentclass{raa_twocolumn}

\usepackage{graphicx,times}
\usepackage{natbib}
\usepackage{amssymb,amsmath}
\bibpunct{(}{)}{;}{a}{}{,}

\usepackage[pagebackref=true]{hyperref}

\begin{document}

\title{Evolution of the radio source position uncertainties in radio astrometric catalogs over the past three decades}

\volnopage{ {\bf 202X} Vol.\ {\bf XX} No. {\bf XX}, 000--000}
\setcounter{page}{1}

\author{Zinovy Malkin \inst{}}
\institute{Pulkovo Observatory, St.~Petersburg, Pulkovskoye Sh. 65, 196140, Russia; {\it zmalkin@zmalkin.com}\\
\vs\no{\small Received 202X Month Day; accepted 202X Month Day}}

\abstract{
In this paper, progress in improving the coordinates uncertainty of extragalactic radio
sources derived from astrometric and geodetic VLBI observations is investigated.
For this purpose, 30 catalogs of radio source positions computed in 1997--2025 were analyzed.
Over these years, the median source position uncertainty represented by the semi-major axis
of the error ellipse for 467 sources in common to all catalogs improved by one order of magnitude:
from 0.156~mas to 0.015~mas.
It was also found that the improvement in the position uncertainty over time follows a power law
with high accuracy.
The number of observations used for determination of the coordinates of the common sources
in the input catalogs increased over time from 1.3 to 14.4~million, also following
a power law.
A discussion of the results led us to the conclusion that the number of observations is
the primary factor in improving the source positions uncertainty compared
to the number of sessions in which the source was observed.
}

\authorrunning{Zinovy Malkin}            %author_head in even pages
\titlerunning{VLBI-derived radio source position uncertainties}  % title_head in odd pages
\maketitle

%%%%%%%%%%%%%%%%%%%%%%%%%%%%%%%%%%%%%%%%%%%%%%%%%%%%%%%%%%%%%%%%%%%%%%%%%%%%%%

\section{Introduction}

Establishing and maintenance of celestial reference frame (CRF)
based on accurate positions of extragalactic radio sources
(radio astrometry) is one of the key tasks solved by the astrometric
and geodetic VLBI from the beginning of its history \citep{Cohen1971}.
The formal error (uncertainty) of these first determinations of the extragalactic
radio source coordinates was at a level of $1-3''$.
However, already by the middle of the 1970s the uncertainty of the source positions
approaching the accuracy of the best (fundamental) optical catalogs of
$\approx0.1''$, and in the 1980s the accuracy of radio astrometric catalogs
reached a level of several mas \citep{Walter1989}.
As the result, the International Astronomical Union (IAU) at its XXIIIrd
General Assembly approved the resolution B2, which recommended that effective
1~January 1998, VLBI-based radio source catalogs will be realizations of
the newly established International Celestial Reference Frame (ICRF).

Since then, the VLBI technique has been constantly developed, and the number
of observations collected in the framework of numerous observing programs
increased manifold, which has led to significant improvement in the accuracy
of the radio astrometric catalogs.
In this paper we analyzed 30 catalogs of radio source positions derived from
astrometric and geodetic VLBI observing programs since 1997
with main purpose to assess the evolution of the uncertainties in the radio source
positions over past 28 years, since 1997 till 2025.

The paper is organized as follows.
Section~\ref{sect:data_analysis} describes the data used in this study,
Section~\ref{sect:position_errors} is devoted to analysis of the source position uncertainties
in the input catalogs, and the last Section sums up the results obtained in this work.

%%%%%%%%%%%%%%%%%%%%%%%%%%%%%%%%%%%%%%%%%%%%%%%%%%%%%%%%%%%%%%%%%%%%%%%%%%%%%%

\section{Data and analysis}
\label{sect:data_analysis}

For this analysis, we used 30 catalogs of radio source positions
computed in 1997--2025 using the Calc/Solve VLBI analysis software
developed at the Goddard Space Flight Center, NASA, and described
in \citet{Ma1986,Ma1998,Fey2015,Charlot2020} and NASA technical
publications referenced therein.
All the catalogs were computed using the same software
and close methodology, albeit with gradual updates over the years.
To compute the source coordinates in these catalogs
VLBI observations at $X$ (7.7--8.8~GHz)
and $S$ (2.2--2.3~GHz) bands were processed.
Observations at $S$ band were only used to compute the ionosphere
correction, source positions in the catalogs are related to $X$ band.

The source position uncertainties in the input catalogs were computed in two modes.
Most of catalogs contain the position uncertainties directly coming
from a global solution using the least square adjustment.
They will be called hereafter catalogs with original uncertainties.
In some cases the original position uncertainties were additionally
inflated to obtain more realistic position errors.
The inflated uncertainties were computed using the following expressions \citep{Fey2015}:
\begin{equation}
\begin{array}{rcl}
\sigma_{\alpha^{\ast}}^{'2} & = & (f_{\alpha}\;\sigma_{\alpha^{\ast}})^{2} + \sigma_{\alpha,0}^2 \,, \\
\sigma_{\delta}^{'2} & = & (f_{\delta}\;\sigma_{\delta}^{'2}) + \sigma_{\delta,0}^2 \,,
\end{array}
\label{eq:error_inftaion}
\end{equation}
where $\sigma_{\alpha^{\ast}}$ and $\sigma_{\delta}$ are original uncertainties
in right ascension multiplied by $\cos\delta$ and declination, respectively,
$\sigma_{\alpha^{\ast}}$ and $\sigma_{\delta}$ are inflated uncertainties,
$f_{\alpha}$  and $f_{\delta}$ are scaling factors,
and $\sigma_{\alpha,0}$ and $\sigma_{\delta,0}$ are error floors.
The scaling factors and error floors for $\alpha$ and $\delta$ are determined
using special procedures, and are arbitrary to some extent.
For example, for the ICRF catalog \citep{Ma1998}
$f_{\alpha} = f_{\delta}$ = 1.5, $\sigma_{\alpha,0} = \sigma_{\delta,0}$ = 0.25~mas,
for the ICRF2 catalog \citep{Fey2015}
$f_{\alpha} = f_{\delta}$ = 1.5, $\sigma_{\alpha,0} = \sigma_{\delta,0}$ = 0.04~mas,
and for the ICRF3-SX catalog \citep{Charlot2020}
$f_{\alpha} = f_{\delta}$ = 1.5, $\sigma_{\alpha,0} = \sigma_{\delta,0}$ = 0.03~mas.

In this work, input catalogs with the original uncertainties were used.
If the catalog is only available in a version with inflated uncertainties,
it was recomputed using the information provided in the original catalog
and inverse expression Eq.~(\ref{eq:error_inftaion}).

\begin{figure}
\centering
\includegraphics[clip,width=\columnwidth]{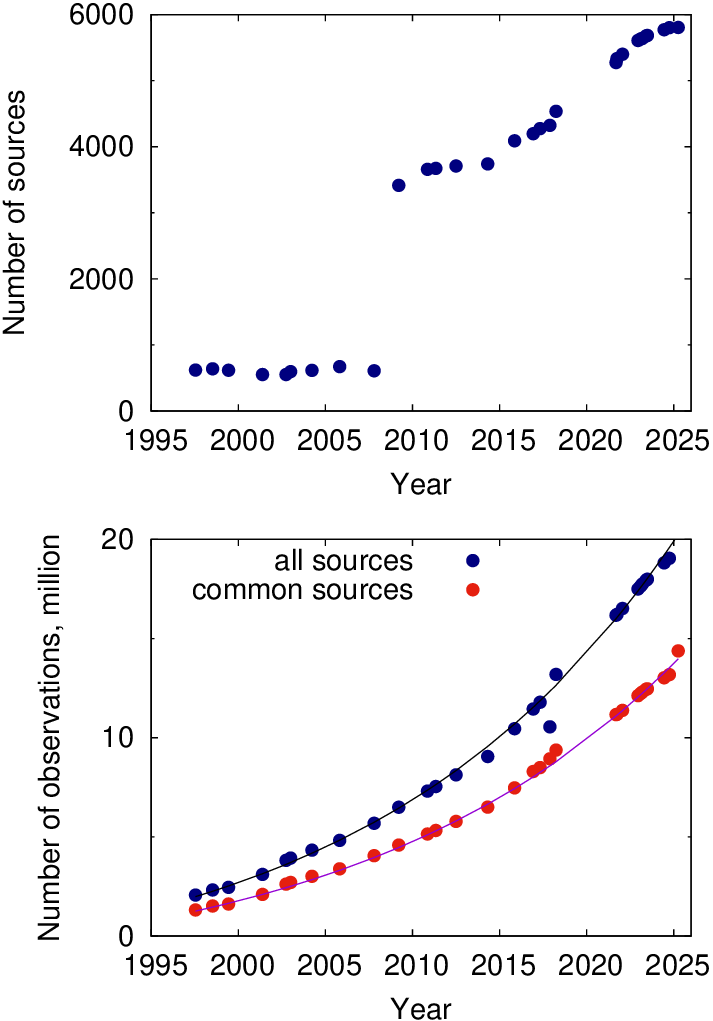}
%includegraphics[clip,width=0.5\columnwidth]{obs_stat.eps}
\caption{Observation statistics for 30 catalogs used in this work:
  the number of sources in each catalog (top panel) and the number of observations
  (bottom panel) for all sources and for 467 sources in common for all catalogs.
  The line in the bottom panel corresponds to a power law approximation of the number
  of observations of common sources.}
\label{fig:obs_stat}
\end{figure}

General information and basic statistics related to the catalogs processed for this study
is presented in Table~\ref{tab:catalogs} and Fig.~\ref{fig:obs_stat}.

\begin{table*}
\centering
\caption{Basic information and statistics for the 30 input catalogs.}
\label{tab:catalogs}
\begin{tabular}{lcc|rrccc|rccc}
\hline
Catalog & \multicolumn{2}{c|}{Last epoch} & \multicolumn{5}{c|}{All sources} & \multicolumn{4}{c}{467 Common sources}\\
& MJD & Year & Nsou & Nobs~~~ & \multicolumn{3}{c|}{Median uncertainty, mas} & Nobs~~~ & \multicolumn{3}{c}{Median uncertainty, mas} \\
&&&&& $\sigma_{\alpha^{\ast}}$ & $\sigma_{\delta}$ & $SMA$ && $\sigma_{\alpha^{\ast}}$ & $\sigma_{\delta}$ & $SMA$ \\
\hline
gsf1997c & 50646 & 1997.54 &  620 &  2063551 & 0.1362 & 0.1817 & 0.1881 &  1314749 & 0.1103 & 0.1461 & 0.1557 \\
gsf1998g & 51003 & 1998.52 &  638 &  2322661 & 0.0973 & 0.1432 & 0.1468 &  1513593 & 0.0828 & 0.1199 & 0.1208 \\
gsf19995 & 51339 & 1999.44 &  617 &  2451068 & 0.0888 & 0.1240 & 0.1300 &  1610531 & 0.0767 & 0.1118 & 0.1137 \\
gsf2001c & 52053 & 2001.39 &  552 &  3103286 & 0.0580 & 0.0865 & 0.0882 &  2096415 & 0.0580 & 0.0866 & 0.0885 \\
gsf2002c & 52545 & 2002.74 &  552 &  3815486 & 0.0474 & 0.0724 & 0.0731 &  2614614 & 0.0470 & 0.0713 & 0.0726 \\
gsf2003a & 52640 & 2003.00 &  596 &  3928397 & 0.0494 & 0.0762 & 0.0778 &  2699910 & 0.0453 & 0.0707 & 0.0712 \\
gsf2004b & 53091 & 2004.23 &  615 &  4331451 & 0.0469 & 0.0723 & 0.0727 &  3011868 & 0.0423 & 0.0648 & 0.0657 \\
gsf2005e & 53672 & 2005.82 &  671 &  4824569 & 0.0426 & 0.0649 & 0.0657 &  3390553 & 0.0359 & 0.0541 & 0.0544 \\
gsf2007d & 54393 & 2007.80 &  609 &  5685251 & 0.0333 & 0.0497 & 0.0497 &  4048831 & 0.0308 & 0.0450 & 0.0452 \\
ICRF2    & 54908 & 2009.21 & 3414 &  6491828 & 0.2632 & 0.4919 & 0.5098 &  4585559 & 0.0271 & 0.0380 & 0.0384 \\
gsf2010a & 55511 & 2010.86 & 3658 &  7309370 & 0.2717 & 0.5155 & 0.5401 &  5143723 & 0.0244 & 0.0340 & 0.0350 \\
gsf2011a & 55686 & 2011.34 & 3671 &  7540393 & 0.2685 & 0.5110 & 0.5326 &  5321900 & 0.0228 & 0.0330 & 0.0337 \\
gsf2012a & 56110 & 2012.50 & 3708 &  8128449 & 0.2389 & 0.4540 & 0.4735 &  5781338 & 0.0201 & 0.0290 & 0.0293 \\
gsf2014a & 56776 & 2014.32 & 3740 &  9049465 & 0.2102 & 0.3985 & 0.4123 &  6498120 & 0.0186 & 0.0260 & 0.0267 \\
gsf2015b & 57337 & 2015.86 & 4089 & 10453527 & 0.1228 & 0.2100 & 0.2181 &  7469980 & 0.0170 & 0.0226 & 0.0228 \\
gsf2016b & 57729 & 2016.93 & 4196 & 11447130 & 0.1286 & 0.2131 & 0.2230 &  8297258 & 0.0166 & 0.0223 & 0.0226 \\
gsf2017a & 57873 & 2017.33 & 4273 & 11791279 & 0.1066 & 0.1835 & 0.1905 &  8493225 & 0.0157 & 0.0214 & 0.0216 \\
gsf2018c & 58080 & 2017.89 & 4323 & 10547496 & 0.0838 & 0.1460 & 0.1488 &  8935897 & 0.0150 & 0.0209 & 0.0211 \\
ICRF3-SX & 58206 & 2018.24 & 4536 & 13190228 & 0.0821 & 0.1439 & 0.1471 &  9379123 & 0.0144 & 0.0201 & 0.0202 \\
usn2021a & 59466 & 2021.69 & 5274 & 16173976 & 0.0678 & 0.1203 & 0.1229 & 11163384 & 0.0139 & 0.0185 & 0.0186 \\
usn2021b & 59486 & 2021.74 & 5333 & 16219604 & 0.0684 & 0.1216 & 0.1240 & 11179607 & 0.0139 & 0.0185 & 0.0186 \\
usn2022a & 59601 & 2022.06 & 5399 & 16514351 & 0.0688 & 0.1218 & 0.1242 & 11381237 & 0.0139 & 0.0183 & 0.0183 \\
usn2022d & 59927 & 2022.95 & 5608 & 17497505 & 0.0696 & 0.1245 & 0.1271 & 12122013 & 0.0135 & 0.0173 & 0.0174 \\
usn2023a & 59983 & 2023.10 & 5629 & 17643510 & 0.0696 & 0.1249 & 0.1277 & 12230629 & 0.0135 & 0.0172 & 0.0173 \\
usn2023b & 60020 & 2023.20 & 5639 & 17750815 & 0.0694 & 0.1247 & 0.1271 & 12306145 & 0.0135 & 0.0171 & 0.0173 \\
usn2023c & 60097 & 2023.41 & 5676 & 17940636 & 0.0693 & 0.1245 & 0.1269 & 12440866 & 0.0133 & 0.0170 & 0.0172 \\
usn2023d & 60123 & 2023.49 & 5683 & 17994850 & 0.0691 & 0.1241 & 0.1267 & 12466656 & 0.0133 & 0.0170 & 0.0172 \\
usn2024a & 60476 & 2024.45 & 5770 & 18819817 & 0.0684 & 0.1244 & 0.1267 & 13024993 & 0.0134 & 0.0170 & 0.0171 \\
usn2024b & 60585 & 2024.75 & 5802 & 19047642 & 0.0690 & 0.1255 & 0.1279 & 13179020 & 0.0134 & 0.0170 & 0.0170 \\
usn2025a & 60770 & 2025.26 & 5804 & 20570840 & 0.0606 & 0.1117 & 0.1137 & 14379983 & 0.0117 & 0.0149 & 0.0149 \\
\hline
\end{tabular}\\
\textbf{Notes.} Nsou is the number of sources in the catalog, Nobs is the number of observations
  processed to compute the catalog, $\sigma_{\alpha^{\ast}}$ is the uncertainty in right ascension
  multiplied by $\cos\delta$, $\sigma_{\delta}$ is the uncertainty in declination, $SMA$ is the semi-major
  axis of the error ellipse.
\end{table*}

\begin{figure}
\centering
\includegraphics[clip,width=\columnwidth]{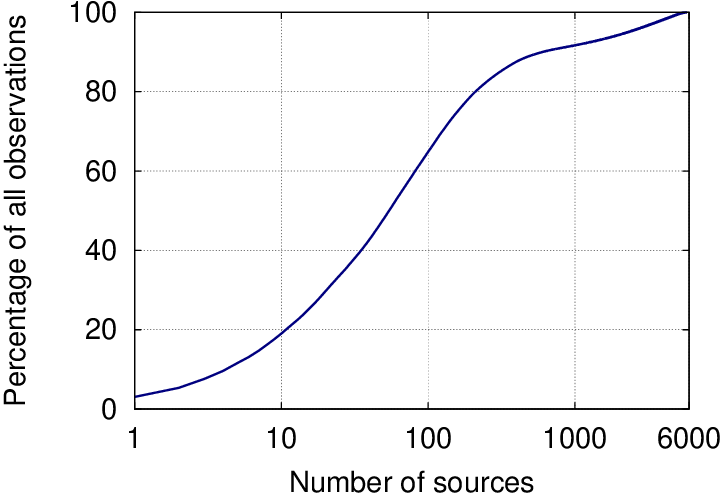}
\caption{Cumulative number of observations depending on the number of sources
  for the usn2025a catalog. Summations starts from sources with the maximum
  number of observations.}
\label{fig:nobs_sources_cumulative}
\end{figure}

The number of observations of the common sources is about 70\% of the total number
of observations.
See also Fig.~\ref{fig:nobs_sources_cumulative} with observation statistics
for the latest catalog used in this study, usn2025a.

Unlike the plot with the number of sources, there is no visible jump in
the plots of the number of observations.
The main reason is that although a large number of sources were added
to the radio astrometric catalogs after 2008, the number of observations of these
new sources is relatively small as compared with ``old'' sources regularly observed
in the VLBI geodetic observing programs.

%%%%%%%%%%%%%%%%%%%%%%%%%%%%%%%%%%%%%%%%%%%%%%%%%%%%%%%%%%%%%%%%%%%%%%%%%%%%%%

\section{Source position uncertainties}
\label{sect:position_errors}

\begin{figure}
\centering
\includegraphics[clip,width=\columnwidth]{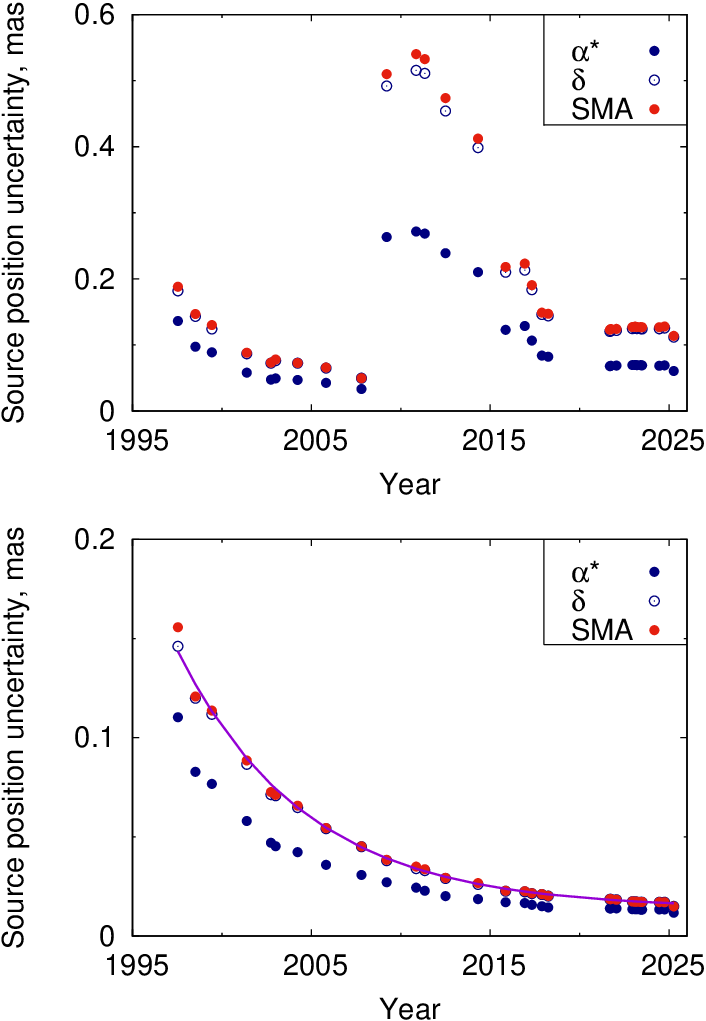}
\caption{Median position uncertainties in $\alpha^{\ast}$ and $\delta$, and the semi-major axis
  of the error ellipse (SMA) for 30 catalogs used in this work for all sources (top panel)
  and 467 common sources (bottom panel). The line in the bottom panel corresponds
  to a power law approximation of the SMA data.}
\label{fig:median_errors}
\end{figure}

Figure~\ref{fig:median_errors} shows the evolution median position
(the semi-major axis of the error ellipse, SMA) over 28 years.
The top plot presents uncertainties for all catalogs including all included sources.
As was mentioned above, abrupt increasing of the number of source after 2009
caused the increase in the median position uncertainty because most new sources
have less accurate positions due to more poor observation history.
The lower plot in Fig.~\ref{fig:median_errors} shows only data for 467 sources
in common for all 30 catalogs, and it can be used to assess the evolution
of the source position uncertainties over time.

Overall improvement in the position uncertainty is about one order of magnitude.
Uncertainty in $\delta$ is greater than uncertainty in $\alpha^{\ast}$ in all
catalogs because of stronger distribution of latitude-oriented baselines,
as is well known from previous analysis.
For this reason, the SMA value is dominated by the uncertainty in $\delta$.

\begin{figure}
\centering
\includegraphics[clip,width=\columnwidth]{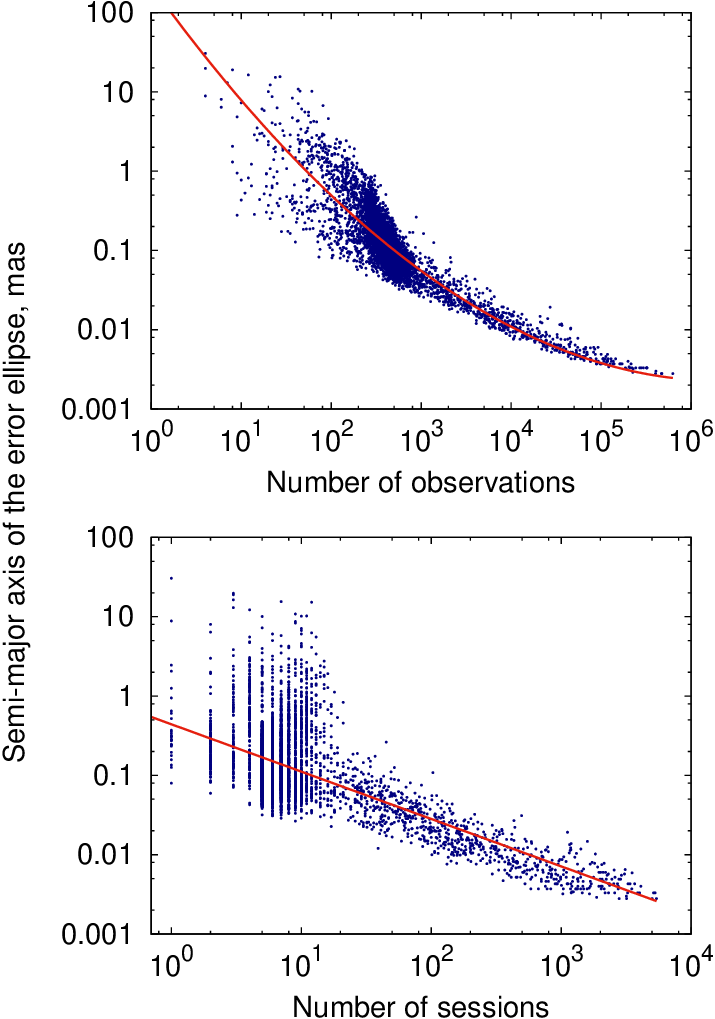}
\caption{Dependence of the source position uncertainty represented by the semi-major
  axis of the error ellipse on the number of observations (top panel) and
  the number of sessions (bottom panel) for the usn2025a catalog.
  One source with the position uncertainty of 141~mas is not shown.}
\label{fig:smaja_obs_stat}
\end{figure}

Comparing Fig.~\ref{fig:median_errors} (top panel) with Fig.~\ref{fig:obs_stat}
(top panel) one can conclude that massive addition of new sources
caused substantial degradation of the median position uncertainties.
The main reason of this is that the additional sources have worse observation history.
It is interesting to estimate how the source position uncertainties depend on the number
of sessions in which the source was observed and the total number of observations used
for computation of the source coordinates.
Figure~\ref{fig:smaja_obs_stat} presented corresponding data for the latest
catalog usn2025a.

It can be concluded from comparison of the scatter of the data points
on the top and bottom panels of Fig.~\ref{fig:smaja_obs_stat}
that the dependence of the uncertainties in the source position
on the observation history is better described by the number of observations
rather than by the number of sessions.
Moreover, in the latter case, such a dependence appears very weak for sources observed
in fewer than about 10 sessions.
In contrast, the source position uncertainty improves more rapidly with increasing
the number of observations for observations less than about 1000 than with further
increases in the number of observations.
The dependence of the SMA on the number of observations ($N$obs) can be approximated
by a quadratic function, while the dependence of the SMA on the number
of sessions ($N$sess) can be approximated by a linear function, as follows:
\begin{equation}
\begin{array}{rcl}
\log\mathrm{SMA} &=& 2.34 - 1.57\log{N\mathrm{obs}} + 0.124\log^2{N\mathrm{obs}} \,, \\
          &=& -0.36 - 0.60\log{N\mathrm{sess}} \,,
\end{array}
\label{eq:sma_nobs}
\end{equation}
where SMA is expressed in mas.
Red lines in Fig.~\ref{fig:smaja_obs_stat} correspond these approximations.

Comparison of Fig.~\ref{fig:obs_stat} and Fig.~\ref{fig:median_errors} also shows
that both increasing of the number of observations of the common sources over time
and decreasing of the median position uncertainty follow a power law.

\begin{figure}
\centering
\includegraphics[clip,width=\columnwidth]{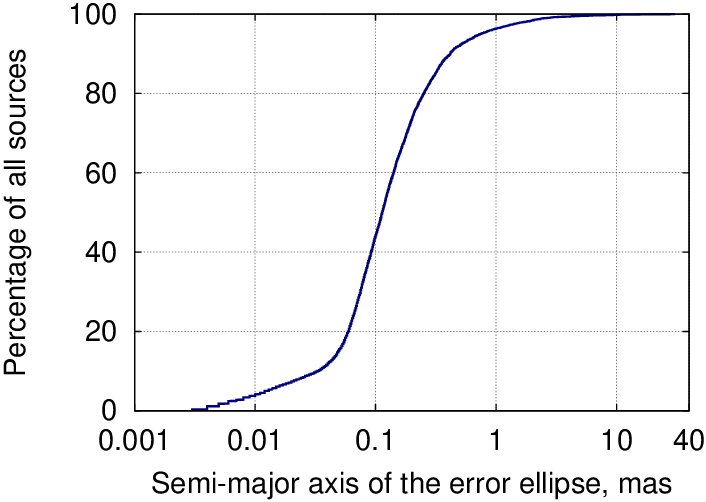}
\caption{Cumulative number of sources with the position uncertainty represented
  by the semi-major axis of the error ellipse is less than the value given
  on the $x$ axis (usn2025a catalog).
  One source with the position uncertainty of 141~mas is not shown.}
\label{fig:smaja_cumulative}
\end{figure}

Figure~\ref{fig:smaja_cumulative} presents the distribution of the source position
uncertainties (SMA) in the usn2025a catalog.
It shows the percentage of sources with the position uncertainty less than the given
value shown on the abscissa axis.
In particular, more than 95\% sources has position uncertainty less than 1~mas, and
about 40\% sources has position uncertainty less than 0.1~mas.
The whole range of the position uncertainties is from 3~$\mu$as to 141~mas
(31~mas without one source).

%%%%%%%%%%%%%%%%%%%%%%%%%%%%%%%%%%%%%%%%%%%%%%%%%%%%%%%%%%%%%%%%%%%%%%%%%%%%%%

\section{Conclusions}
\label{sect:conclusions}

An interesting fact was noticed that both the increase in the number of observations
used to derive the source coordinates and the decrease in the median source position
uncertainties follow a power law with high accuracy.
Of course, an increase in the number of observations may be not the only factor
in improving the source position uncertainties over time.
Improving the VLBI technology, extending the VLBI networks, and developings in the data
analysis should also play a role in improving the accuracy of catalogs over time.
However, it is difficult, if possible at all, to quantify the contribution of these
factors because the sources were observed in different sessions of different design
and not simultaneously.
On the other hand, it can be expected that these improvements in the VLBI technique
should manifest themselves in improved catalogs of radio source position.
As was shown by \citet{Malkin2020}, the accuracy of the baseline length determinations
has only slightly improved after the late 1900s, and the accuracy of the Earth
orientation parameters, although has improved after the late 1900s, but to a much
lesser extent compared with the improvement in the accuracy of source positions.
This allows as to suggest that the number of observations is the primary factor
determining the accuracy of the VLBI-based radio source catalogs.

Generally speaking, it can be noticed that the power law can often serve as a means
of describing various features of astrometric and geodetic VLBI observations.
For example, in addition to the results of the present study, it was shown
in \citet{Malkin2009} that both precision and accuracy of the EOP obtained
from VLBI observations can be well described by a power function
of the geometric volume of the network.

It was also shown that the dependence of the position uncertainties on the source
observation history is better described by the number of observations rather
than by the number of sessions, which can be taken into account for optimal
scheduling of the astrometric VLBI observations.
As follows from this result, observing fewer sessions with more observations
appears to be preferable to observing more sessions with a smaller total
number of observations.
The former strategy is also preferable for better source imaging.

%%%%%%%%%%%%%%%%%%%%%%%%%%%%%%%%%%%%%%%%%%%%%%%%%%%%%%%%%%%%%%%%%%%%%%%%%%%%%%

\section*{Acknowledgments}

The author would like to express his gratitude to the editor and reviewer who took
the time to review this submission and for their valuable comments and suggestions.
The author is very grateful to his colleagues from the VLBI analysis centers
at the NASA Goddard Space Flight Center and U.~S. Naval Observatory, especially
David Gordon, who provided their radio source positions catalogs used in this study.
This research has made use of SAO/NASA Astrophysics Data
System\footnote{\url{https://ui.adsabs.harvard.edu/}} (ADS).
Figures were prepared using \texttt{gnuplot}\footnote{\url{http://www.gnuplot.info/}}.

\bibliography{catalog_errors}
\bibliographystyle{raa}

\end{document}